\documentclass[twoside,twocolumn,9pt]{article}
\usepackage{extsizes}
\usepackage[super,sort&compress,comma]{natbib}
\usepackage[version=3]{mhchem}
\usepackage[left=1.5cm, right=1.5cm, top=1.785cm, bottom=2.0cm]{geometry}
\usepackage{balance}
\usepackage{widetext}
\usepackage{times,mathptmx}
\usepackage{sectsty}
\usepackage{graphicx}
\usepackage{lastpage}
\usepackage[format=plain,justification=raggedright,singlelinecheck=false,font={stretch=1.125,small,sf},labelfont=bf,labelsep=space]{caption}
\usepackage{float}
\usepackage{fancyhdr}
\usepackage{fnpos}
\usepackage[english]{babel}
\usepackage{array}
\usepackage{droidsans}
\usepackage{charter}
\usepackage[T1]{fontenc}
\usepackage[usenames,dvipsnames]{xcolor}
\usepackage{setspace}
\usepackage[compact]{titlesec}

\definecolor{cream}{RGB}{222,217,201}

\begin{document}

\pagestyle{fancy}
\thispagestyle{plain}
\fancypagestyle{plain}{

\fancyhead[C]{\includegraphics[width=18.5cm]{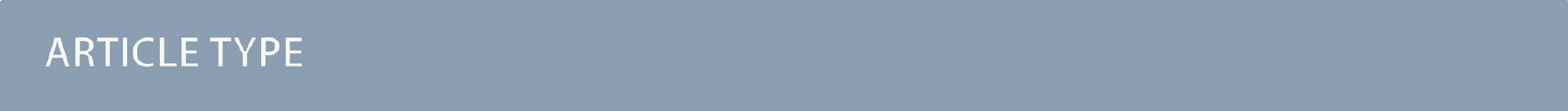}}
\fancyhead[L]{\hspace{0cm}\vspace{1.5cm}\includegraphics[height=30pt]{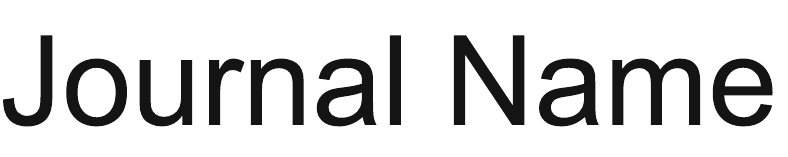}}
\fancyhead[R]{\hspace{0cm}\vspace{1.7cm}\includegraphics[height=55pt]{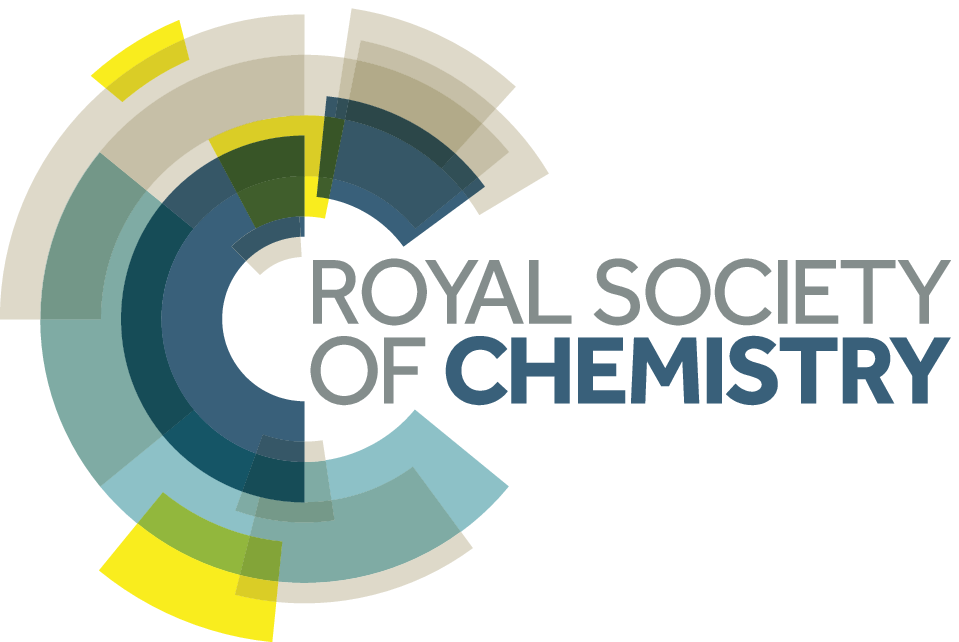}}
\renewcommand{\headrulewidth}{0pt}
}

\makeFNbottom
\makeatletter
\renewcommand\LARGE{\@setfontsize\LARGE{15pt}{17}}
\renewcommand\Large{\@setfontsize\Large{12pt}{14}}
\renewcommand\large{\@setfontsize\large{10pt}{12}}
\renewcommand\footnotesize{\@setfontsize\footnotesize{7pt}{10}}
\makeatother

\renewcommand{\thefootnote}{\fnsymbol{footnote}}
\renewcommand\footnoterule{\vspace*{1pt}%
\color{cream}\hrule width 3.5in height 0.4pt \color{black}\vspace*{5pt}}
\setcounter{secnumdepth}{5}

\makeatletter
\renewcommand\@biblabel[1]{#1}
\renewcommand\@makefntext[1]%
{\noindent\makebox[0pt][r]{\@thefnmark\,}#1}
\makeatother
\renewcommand{\figurename}{\small{Fig.}~}
\sectionfont{\sffamily\Large}
\subsectionfont{\normalsize}
\subsubsectionfont{\bf}
\setstretch{1.125} 
\setlength{\skip\footins}{0.8cm}
\setlength{\footnotesep}{0.25cm}
\setlength{\jot}{10pt}
\titlespacing*{\section}{0pt}{4pt}{4pt}
\titlespacing*{\subsection}{0pt}{15pt}{1pt}

\fancyfoot{}
\fancyfoot[LO,RE]{\vspace{-7.1pt}\includegraphics[height=9pt]{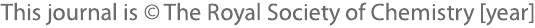}}
\fancyfoot[CO]{\vspace{-7.1pt}\hspace{13.2cm}\includegraphics{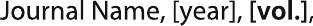}}
\fancyfoot[CE]{\vspace{-7.2pt}\hspace{-14.2cm}\includegraphics{head_foot/RF}}
\fancyfoot[RO]{\footnotesize{\sffamily{1--\pageref{LastPage} ~\textbar  \hspace{2pt}\thepage}}}
\fancyfoot[LE]{\footnotesize{\sffamily{\thepage~\textbar\hspace{3.45cm} 1--\pageref{LastPage}}}}
\fancyhead{}
\renewcommand{\headrulewidth}{0pt}
\renewcommand{\footrulewidth}{0pt}
\setlength{\arrayrulewidth}{1pt}
\setlength{\columnsep}{6.5mm}
\setlength\bibsep{1pt}

\makeatletter
\newlength{\figrulesep}
\setlength{\figrulesep}{0.5\textfloatsep}

\newcommand{\topfigrule}{\vspace*{-1pt}%
\noindent{\color{cream}\rule[-\figrulesep]{\columnwidth}{1.5pt}} }

\newcommand{\botfigrule}{\vspace*{-2pt}%
\noindent{\color{cream}\rule[\figrulesep]{\columnwidth}{1.5pt}} }

\newcommand{\dblfigrule}{\vspace*{-1pt}%
\noindent{\color{cream}\rule[-\figrulesep]{\textwidth}{1.5pt}} }

\makeatother

\twocolumn[
  \begin{@twocolumnfalse}
\vspace{3cm}
\sffamily
\begin{tabular}{m{4.5cm} p{13.5cm} }

\includegraphics{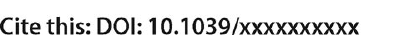} & \noindent\LARGE{\textbf{Determination of the exciton binding energy and effective masses for the methylammonium and formamidinium lead tri-halide perovskite family}} \\
\vspace{0.3cm} & \vspace{0.3cm} \\

 & \noindent\large{Krzysztof Galkowski,$^{\ast}$\textit{$^{a,b}$} Anatolie Mitioglu,\textit{$^{a}$} Atsuhiko
 Miyata,\textit{$^{a}$} Paulina Plochocka,\textit{$^{a}$} Oliver
 Portugall,\textit{$^{a}$} Giles E. Eperon,\textit{$^{c}$} Jacob Tse-Wei Wang,\textit{$^{c}$} Thomas Stergiopoulos,\textit{$^{c}$} Samuel D. Stranks,\textit{$^{c}$}
 Henry J. Snaith,\textit{$^{c}$} and Robin J. Nicholas\textit{$^{c}$}} \\

\includegraphics{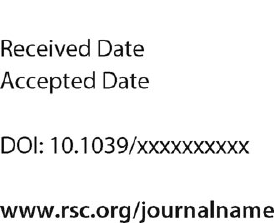} & \noindent\normalsize{The family of organic-inorganic halide perovskite materials has generated
tremendous interest in the field of photovoltaics due to their
high power conversion efficiencies. There has been intensive
development of cells based on the archetypal methylammonium (MA)
and recently introduced formamidinium (FA) materials, however,
there is still considerable controversy over their fundamental
electronic properties. Two of the most important parameters are
the binding energy of the exciton (R$^{*}$) and its reduced
effective mass $\mu$. Here we present extensive  magneto optical
studies of Cl assisted grown MAPbI$_{3}$ as well as MAPbBr$_{3}$
and the FA based materials FAPbI$_{3}$ and FAPbBr$_{3}$. We fit
the excitonic states as a hydrogenic atom in magnetic field and
the Landau levels for free carriers to give R$^{*}$ and $\mu$. The
values of the exciton binding energy are in the range 14 - 25 meV
in the low temperature phase and fall considerably at higher
temperatures for the tri-iodides, consistent with free carrier
behaviour in all devices made from these materials. Both R$^{*}$
and $\mu$ increase approximately proportionally to the band gap,
and the mass values, 0.09-0.117 m$_{0}$, are consistent with a
simple \textbf{k.p} perturbation approach to the band structure
which can be generalized to predict values for the effective mass
and binding energy for other members of this perovskite family of materials.} \\
\end{tabular}

 \end{@twocolumnfalse} \vspace{0.6cm}

  ]


\renewcommand*\rmdefault{bch}\normalfont\upshape
\rmfamily
\section*{}
\vspace{-1cm}


\footnotetext{\textit{$^{a}$~Laboratoire National des Champs
Magnetiques Intenses, CNRS-UJF-UPS-INSA, 143 Avenue de Rangueil,
31400 Toulouse, France}}

\footnotetext{\textit{$^{b}$~Institute of Experimental Physics,
Faculty of Physics, University of Warsaw - Pasteura 5, 02-093
Warsaw, Poland}}

\footnotetext{\textit{$^{c}$~University of Oxford, Clarendon
Laboratory, Parks Road, Oxford, OX1 3PU, United Kingdom; E-mail:
r.nicholas@physics.ox.ac.uk}}





\section{Broader context}
The recent development of organic/inorganic perovskite
semiconductors has had a dramatic impact on the field of thin-film
solar cells leading to efficiencies of over 20$\%$ in materials
such as MAPbI$_{3}$. We have recently shown that significant
factors contributing to their remarkable performance are the small
excitonic reduced effective mass and the small exciton binding
energy. Expanding the perovskite family to materials with a range
of different band gaps is opening up the potential of this
materials system for a range of different applications, including
the design of tandem PV cells and other optoelectronic components
such as lasers and light emitting diodes.  Knowledge of basic
materials parameters such as the effective masses of the charge
carriers is vital to the use and design of devices using
perovskites. Here we show that the effective masses remain small
in this system up to band gaps as high as 2.3 eV, as found in
MAPbBr$_{3}$ and the exciton binding energy in this system is also
much smaller than previously reported. This suggests that
sophisticated devices using large band gap perovskites can be
expected to show free carrier behaviour at room temperature and
will make device design significantly easier. Exchanging the
organic cations (Methyammonium or Formamodimium) is also shown to
have very little effect on their bandstructure, suggesting that
they can be used interchangeably to enhance device performance and
stability.

\section{Introduction}

Organic-inorganic hybrid materials, in particular the tri-halide
perovskites, have been driving a rapid series of breakthroughs in
the field of photovoltaic (PV) devices with conversion
efficiencies already up to 20$\%$ \cite{Zhou14} and beyond in
devices which can be produced on cheap non-crystalline substrates
by both vapour deposition \cite{Liu13}and solution
processing\cite{Lee12,Burschka13}.Their excellent performance
originates from the combination of strong
absorption\cite{Tanaka03} and long diffusion
lengths\cite{Xing13,Stranks13}. In addition, the ability to
generate a family of materials with band gaps which can be tuned
by alloying the halide elements\cite{Noh13} and modifying the
organic ions\cite{Eperon14}, allows the band structure to be
optimized and offers the possibility of their use in
multi-junction cells with even higher efficiencies. In addition to
PV devices the perovskite family of semiconductors has also been
used to successfully demonstrate light emitting
diodes\cite{Tan14}, lasers\cite{Deschler14,Saliba15} and
photodetectors\cite{Stranks15}. The chemical structure of these
perovskite semiconductors is ABX$_{3}$ where
A=CH$_{3}$NH$^{+}_{3}$ = MA (MethylAmmonium) or
A=CH(NH$_{2}$)$_{2}$ = FA (FormAmidinium), B=Pb$^{2+}$; and X =
Cl$^{-}$, I$^{-}$ or Br$^{-}$, or an alloyed combination of
these), where the majority of effort has gone into the development
of the MA family\cite{Kojima09,Lotsch14,Snaith13}. Although there
have been tremendous strides made in developing working PV devices
using these perovskites
\cite{Lee12,Burschka13,Jeng13,Liu14,Liu13,Chen14,Eperon14,Liang14,Wang14}
there have been far fewer studies reported on their fundamental
electronic properties. Basic physical parameters such as the
carrier effective masses ($\mu$) or the exciton binding energy
($R^{*}$), remain poorly characterized and their values are still
controversial. Knowledge of these parameters is crucial for
optimization and control of solar cells based on the
organic-inorganic tri-halide perovskite family.

\begin{figure}[]
\begin{center}
\includegraphics[width=1.0\columnwidth]{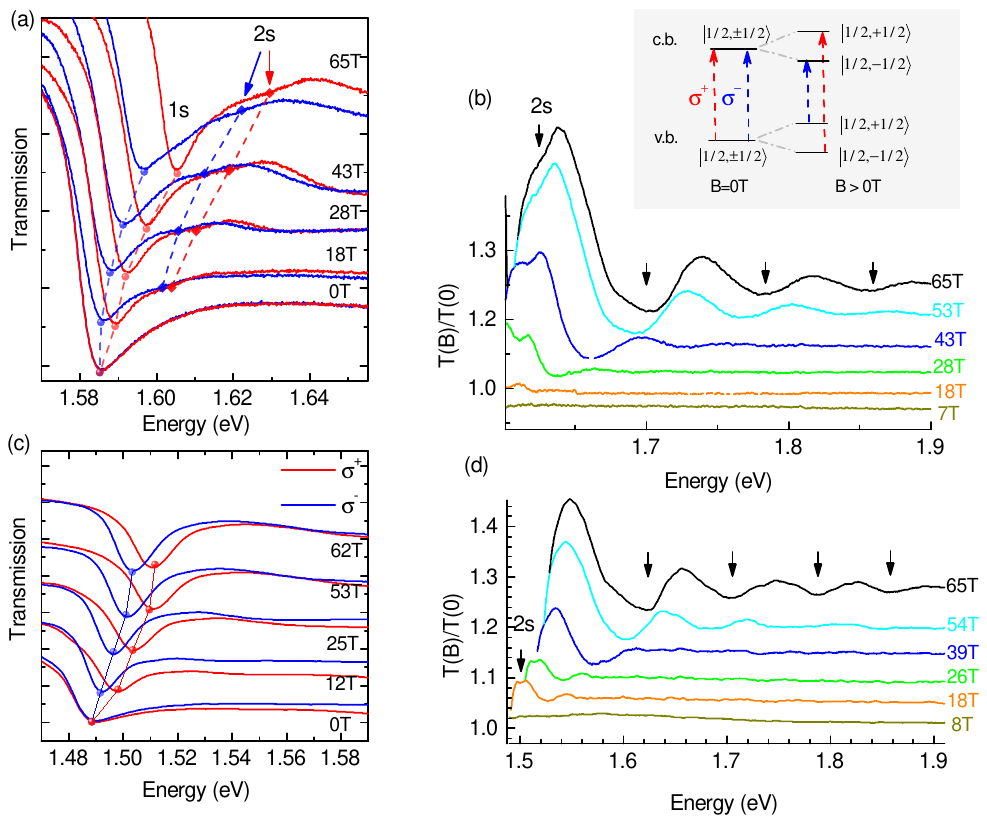}
\end{center}
\caption{(a)-(b) and (c),(d) typical low temperature transmission
data taken using a long pulse technique for MAPbI$_{3-x}$Cl$_{x}$
and FAPbI$_{3}$ respectively. Red and blue lines in panels (a) and
(c) correspond to two $\sigma^{+}$ and $\sigma^{-}$ polarized
light respectively. The selection rules for both polarizations are
shown in the inset. Different colors on panels (b) and (d)
correspond to the data taken at different values of magnetic
field. }\label{Fig1}
\end{figure}

From the point of view of applications the binding energy of the
exciton is critical for determining the operation of devices. This
value compared with the thermal energy (kT $\simeq$ 25 meV at
300K) determines whether the photo created carriers will
dissociate or will need to be separated by an additional junction
in a PV device. Until very recently the reported values of binding
energy were highly controversial and covered a wide range: For
MAPbI$_3$ values from 2 to 55 meV
\cite{Tanaka03,Incenzo14,Hirasawa94,Menendez14,Even14,Yamada15,Lin15}
have been reported with values at the lower end of the range now
becoming more common in the
literature\cite{Even14,Yamada15,Lin15}; and for MAPbBr$_3$ it has
been suggested that $R^{*}$ could be as high as 70
meV\cite{Tanaka03}. Recently, we \cite{Miyata15} have shown that
by using magneto optical studies an exact determination of the
binding energy of the exciton and effective mass of the carriers
for MAPbI$_{3}$ is possible at low temperatures from the
measurement of sequences of higher states for the excitonic
transitions and the quantization of the free carrier states into
Landau levels. We now extend this type of study to the wider
family of organic-inorganic perovskite tri-halides to establish
their basic materials parameters and enable their use in the
design of the new generation of opto-electronic components based
on these materials. In particular we report studies of the
FA-based analogues, which appear to be more thermally stable than
their MA counterparts \cite{Eperon14,Stranks15} and which have
been mixed with MA to give the highest device performance to
date\cite{Yang15}.

Here, we present low temperature studies of magneto optical
absorption spectroscopy, which directly probes basic electronic
parameters such as the binding energy of the neutral free exciton
and the excitonic reduced effective mass ($\mu$) of the carriers
for the different families of hybrid inorganic-organic perovskite
materials. We have investigated the methylammonium and
formamidinium lead halide family with band gaps in the range 1500
- 2233 meV. The value of both the exciton binding energy and the
reduced effective mass increase approximately proportionally to
the band gap and their values are consistent with a simple two
band \textbf{k.p} perturbation theory approach to the band
structure which can be generalized to predict values for the
effective mass and binding energy for other members of this
perovskite family of materials.

\section{Results and discussion}

Typical low temperature magneto transmission spectra measured at
2K for thin films ($\approx$350nm thick) deposited onto glass are
shown for representatives of the two perovskite families in
Fig~\ref{Fig1} at a series of magnetic fields. The panels (a)-(b)
show the results for MAPbI$_{3-x}$Cl$_{x}$ and (c) - (d) results
for FAPbI$_{3}$. The spectra shown in Fig~\ref{Fig1}(a) and (c)
exhibit one very strong minimum, which corresponds to the
strongest absorption originating from the 1s state of the neutral
free exciton. Data taken for different circular polarizations (red
and blue lines) clearly show Zeeman splitting, according to the
selection rules for circularly polarized absorption presented in
the inset in
Fig~\ref{Fig1}(b)\cite{Umebayashi03,Kim14,Tanaka03,Even14}, which
shows the spin split electron and hole states of the conduction
and valence band. Also visible in Fig~\ref{Fig1}(a) is the 2s
state which appears as a further weak absorption. To observe
absorption from higher electronic states the measurements have
been extended to much higher energies. As the absorption to the
higher states is weaker we plot the data as differentials; the
spectrum in high magnetic field is divided by the spectrum at zero
magnetic field (Fig~\ref{Fig1}(b),(d)) to show the additional
field induced absorptions. Here we show only one circular
polarization, as the spin splitting is not resolved on such a
large energy scale. In the differential spectra we observe the 2s
state more clearly (Fig~\ref{Fig1}(b),(d)) and a sequence of
additional minima marked by arrows. The absorption minima become
more pronounced with the increase of the magnetic field and their
energy blue shifts.

\begin{figure*}[]
\begin{center}
\includegraphics[width=1.7\columnwidth]{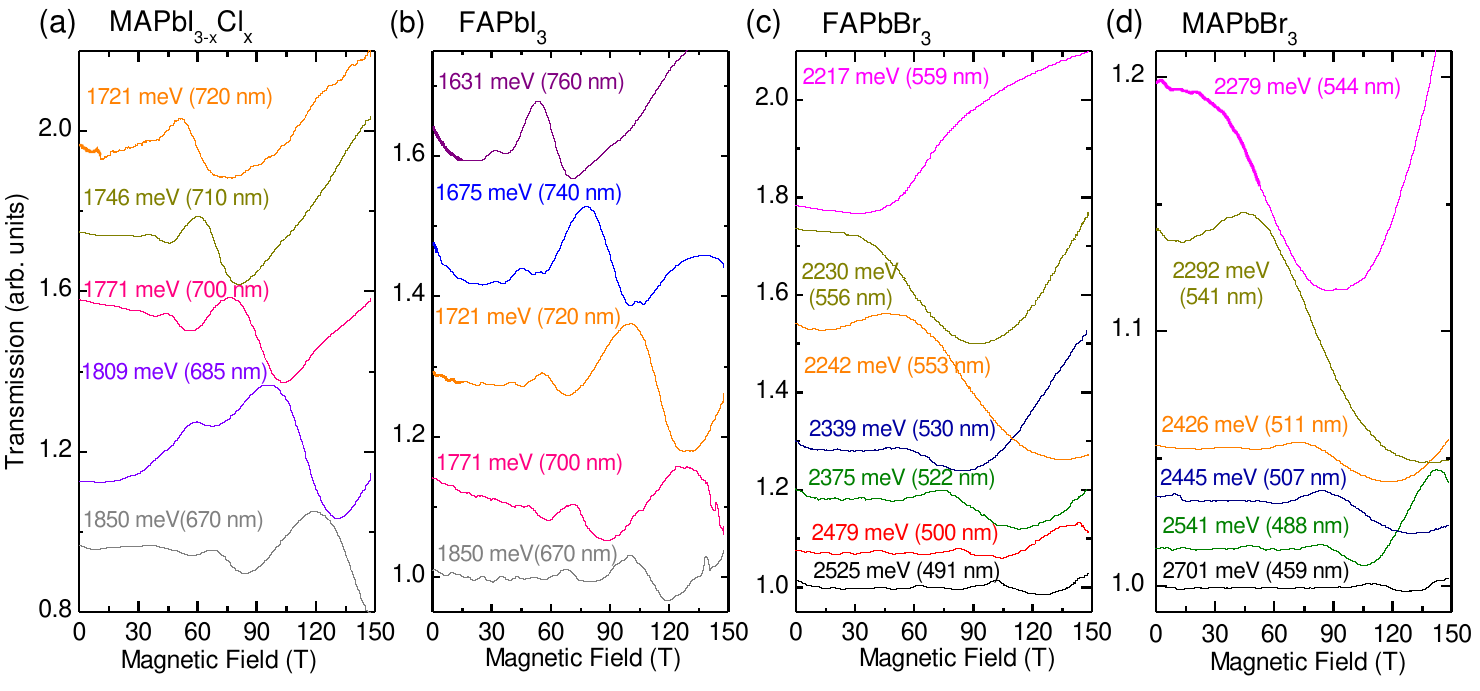}
\end{center}
\caption{(a) - (d) typical results of the low temperature
monochromatic transmission as a function of magnetic field
obtained by using the short pulse technique for
MAPbI$_{3-x}$Cl$_{x}$, FAPbI$_{3}$, FAPbBr$_{3}$ and MAPbBr$_{3}$
respectively. The transmission signals are normalized to 1 at zero
magnetic field and offset to improve visibility. Different colors
mark different laser energies.}\label{Fig2}
\end{figure*}

\begin{figure}[]
\begin{center}
\includegraphics[width=0.8\columnwidth]{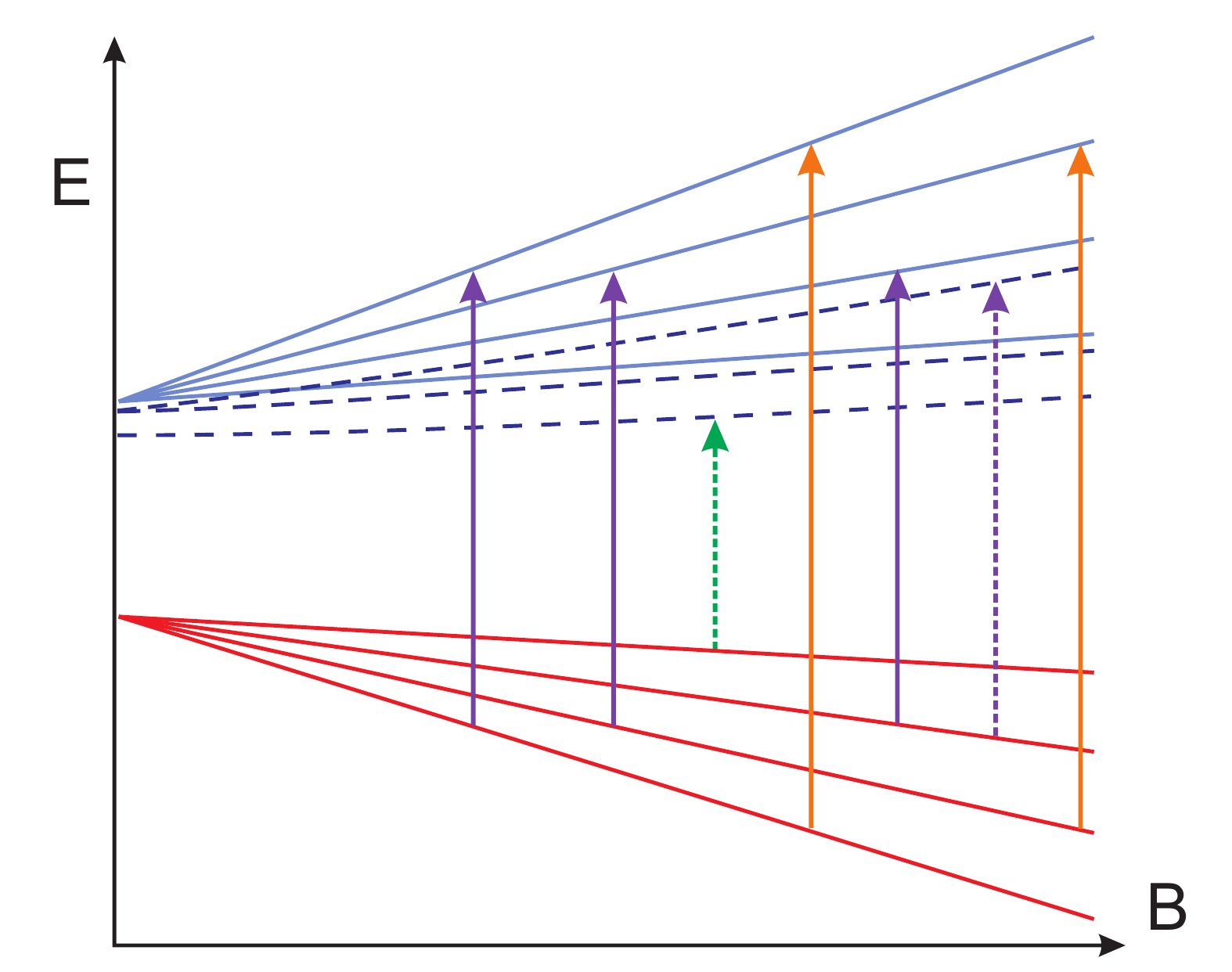}
\end{center}
\caption{A schematic view of the Landau levels (solid lines) and
excitonic states (dashed lines) as a function of magnetic field
showing a sequence of transitions which might be observed for 3
different laser energies}\label{FigIB}
\end{figure}

To follow the absorption minima to higher energies, we extended
the measurements to higher magnetic field (B>70T) using a short
pulse technique where we measure the transmission of a
monochromatic (laser) source through the sample as a function of
magnetic field. Typical results obtained for three different
samples are presented in Fig~\ref{Fig2}. In all the spectra we
observe well developed minima as a function of magnetic field.
This technique gives the equivalent of a horizontal slice of the
absorption spectrum as a function of magnetic field and has
transmission minima at magnetic fields corresponding to resonant
transitions between the free carrier Landau levels of the
conduction and valence band and also for the creation of excitonic
states as shown schematically in Fig~\ref{FigIB}. It is very
powerful for high magnetic fields and high energies where the
transition energies are a strong function of magnetic field and
the level broadening is larger. This is particularly important for
compounds such as FAPbBr$_{3}$, where the band gap is at a rather
high energy (around 2232 meV) and the effective mass is larger,
making the results from the short pulse measurements critical to
determine the correct value for the effective mass of the free
exciton.

\begin{figure*}[]
\begin{center}
\includegraphics[width=1.7\columnwidth]{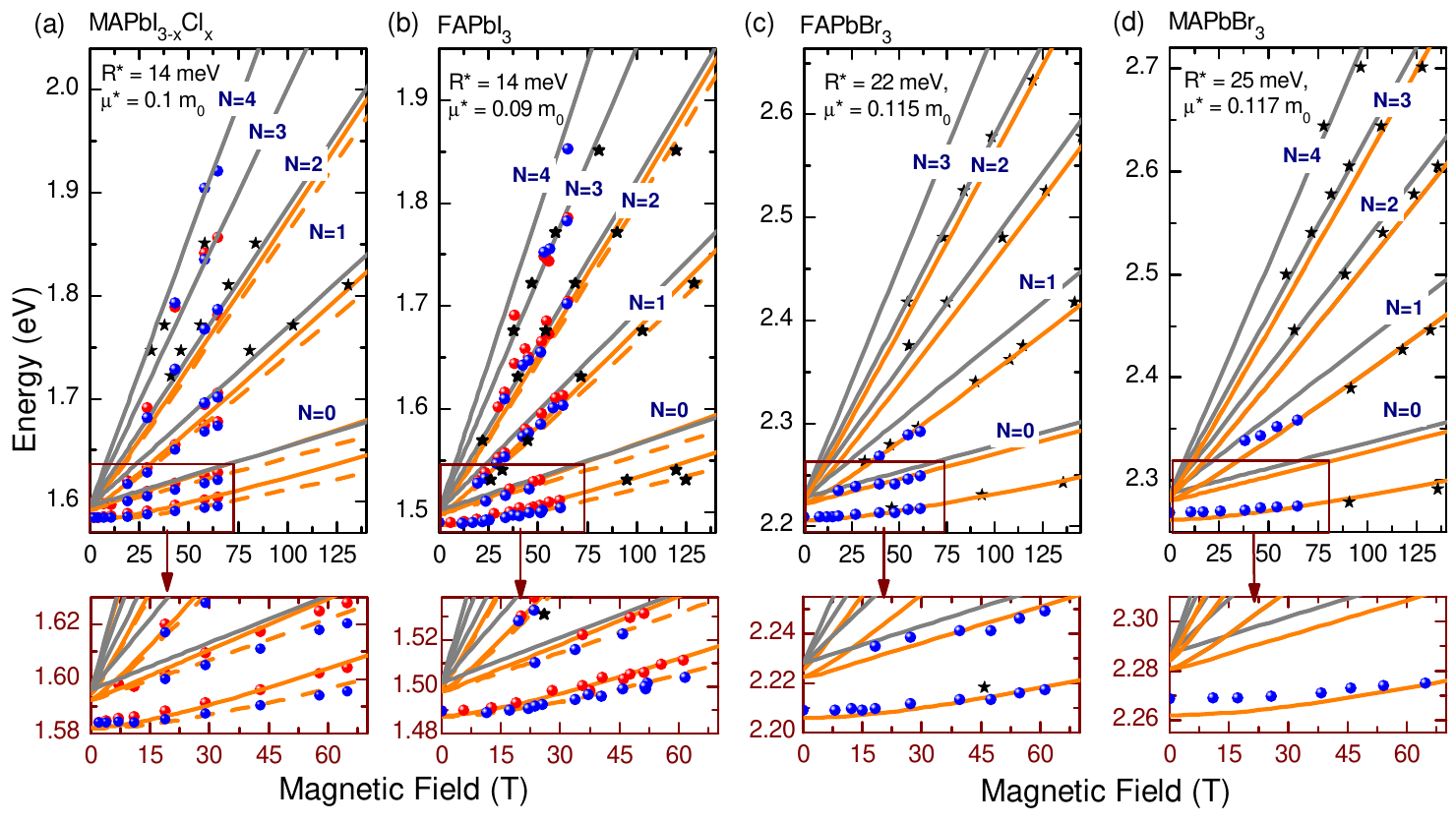}
\end{center}
\caption{(a)-(d) transition energies obtained from the
experimental data and results of the fit to the data for
MAPbI$_{3-x}$Cl$_{x}$, FAPbI$_{3}$, FAPbBr$_{3}$ and MAPbBr$_{3}$
respectively. The red and blue points correspond to the
polarization resolved minima in the absorption for $\sigma^{+}$
and $\sigma^{-}$ respectively. The stars are the data obtained
during the short pulsed measurements, where polarization was not
used. The long pulsed data are also not polarization resolved for
the FAPbBr$_{3}$ and MAPbBr$_{3}$ samples. The results of the
theoretical fit are shown by grey and orange lines. Grey lines
correspond to the interband transitions between Landau levels in
the valence and conduction bands. The orange lines show the
strongly bound levels of the hydrogen - like exciton. The Zeeman
splitting is marked by solid/dashed lines. Below each graph the
low field and low energy part of the full fan chart diagram is
enlarged.}\label{Fig3}
\end{figure*}

In order to extract the binding energy of the free exciton and the
effective mass of the carriers we have analyzed the position of
the absorption energy as a function of magnetic field. The results
of this analysis are presented as fan diagrams in Fig~\ref{Fig3}.
The circles and stars correspond to the minima observed in the
long pulse spectra and the fast pulse magneto-transmission
respectively. To fit the data we have used the same theoretical
approach as previously used~\cite{Miyata15} for MAPbI$_{3}$: (i)
the absorption close to the energy gap is assumed to originate
from hydrogen like behavior typical for free excitons and this
behaviour is stronger in the tri-bromide samples where the exciton
binding energy is found to be larger, (ii) the higher energy part
of the spectrum reflects the free carrier behavior: interband
transitions between Landau levels in the valence and conduction
bands\cite{Watanabe03}. The hydrogen-like behavior can be modelled
using the numerical solutions of the hydrogen atom in a strong
magnetic field calculated by Makado and McGill\cite{Makado19}. The
energy of the free exciton transitions E$_{n,0}(\gamma)$ can be
described using only the dimensionless parameter $\gamma =
\hbar\omega_{c}/2R^{*}$, where $\hbar\omega_{c} = \hbar eB / \mu$
is a cyclotron energy directly proportional to the magnetic field
and inversely proportional to the reduced exciton effective mass
$\mu$ defined by $1/\mu = 1/m_{h}+1/m_{e}$, where $m_{h,e}$ are
the effective masses of the hole and electron respectively. For
high magnetic fields where $\gamma>1$ the free carrier inter
Landau level transitions become dominant with transition energies
given by

\begin{equation}\label{Landaulevels}
    E(B) = E_{g}+(n+\frac{1}{2})\hbar\omega_{c}\pm \frac{1}{2}g_{eff}\mu_{B}B
\end{equation}
where $E_{g}$ is the energy band gap, n = 0,1,2... is a Landau
Level quantum number which is conserved in the transition between
hole and electron levels, $g_{eff}$ is an effective g factor for
the Zeeman splitting and $\mu_{B}$ is the Bohr magneton. As the
Zeeman effect only introduces a constant splitting in the
absorbtion spectra it can be introduced in a similar way into the
hydrogen like model for the excitons. It is important to note that
the Zeeman splitting was not observed in our previous
studies\cite{Miyata15} as circular polarized light was not used.

The results of the fitting for each sample are shown in
Fig~\ref{Fig3} by solid and dashed lines for the excitonic states
in the two polarizations. Transitions between Landau levels are
marked by grey lines. In the case of interband transitions the
only fitting parameter is the reduced mass of the exciton. As
there is only one fitting parameter the high magnetic field and
high energy data give a strong constraint on the exciton reduced
mass. By keeping the same effective mass we can then fit the
hydrogen - like transitions using the exciton binding energy
(R$^{*}$) as another fitting parameter. Moreover, observation of
both the 1s state and in particular the 2s state, which is only
observable once its oscillator strength has been enhanced by the
magnetic field, is very important. Both of these levels originate
from the same Landau level giving a very strong constraint in the
determination of the exciton binding energy, which scales for a 3D
hydrogen atom in zero magnetic field as:

\begin{equation}\label{1s}
    E_{n} = E_{g}-\frac{R^{*}}{n^{2}}
\end{equation}

where $E_{n}$ is the energy of n-th excitonic level and $R^*$ =
$R_0 \mu/m_0\epsilon_{r}^{2}$, where $R_0$ is the atomic Rydberg
constant, $m_0$ is the free electron mass and $\epsilon_{r}$ is
the relative dielectric constant. Results of the analysis for
different compounds are summarized in table~\ref{Tab_t1}.

\begin{table}[h]
2K, Orthorhombic phase
\begin{center}
\begin{tabular}{|l|c|c|c|c|c|}
\hline
 Compound & $E_g$ &$R^{*}$& $\mu$ & $\epsilon_{eff}$ & $g_{eff}$\\
 & (meV) & (meV) & ($m_{e}$) &  & \\
\hline
FAPbI$_{3}$  & 1501 & 14 & 0.09 & 9.35 & 2.3 \\
MAPbI$_{3-x}$Cl$_{x}$ & 1596 & 14 & 0.10 & 9.85 & 2.3 \\
MAPbI$_{3}$ & 1652& 16 & 0.104 & 9.4 & \\
FAPbBr$_{3}$ & 2233& 22 & 0.115 & 8.42 & \\
MAPbBr$_{3}^{*}$ & 2292& 25 & 0.117 & 7.5 & \\
\hline
\end{tabular}
\end{center}
High Temperature Tetragonal phase
\begin{center}
\begin{tabular}{|l|c|c|c|c|c|}
\hline
 Compound & $E_g$ & $R^{*}$ & $\mu$  & $\epsilon_{eff}$ & $Temperature$\\
 & (meV) & (meV) & ($m_{e}$) &  & (K)\\
\hline
FAPbI$_{3}$ & 1521 & 10 & 0.095 & 11.4 & 140-160 \\
MAPbI$_{3-x}$Cl$_{x}$ & 1600 & 10 & 0.105 & 11.9 & 190-200\\
MAPbI$_{3}$  & 1608 & 12 & 0.104 & 10.9 & 155-190 \\
FAPbBr$_{3}$ & 2294 & 24 & 0.13 & 8.6 & 160-170\\
\hline
\end{tabular}
\end{center}
\caption{The parameters of the fit full Landau fan chart for four
different compounds in the low temperature, orthorhombic and the
higher temperature, tetragonal phases. $^{*}$The data for the
MAPbBr$_{3}$ do not show a detectable 2s state and therefore have
a significant errors of $\pm 10\%$ for $\mu$ and $\pm 20\%$ for
$R^{*}$.}\label{Tab_t1}
\end{table}

\begin{figure}[]
\begin{center}
\includegraphics[width=0.9\columnwidth]{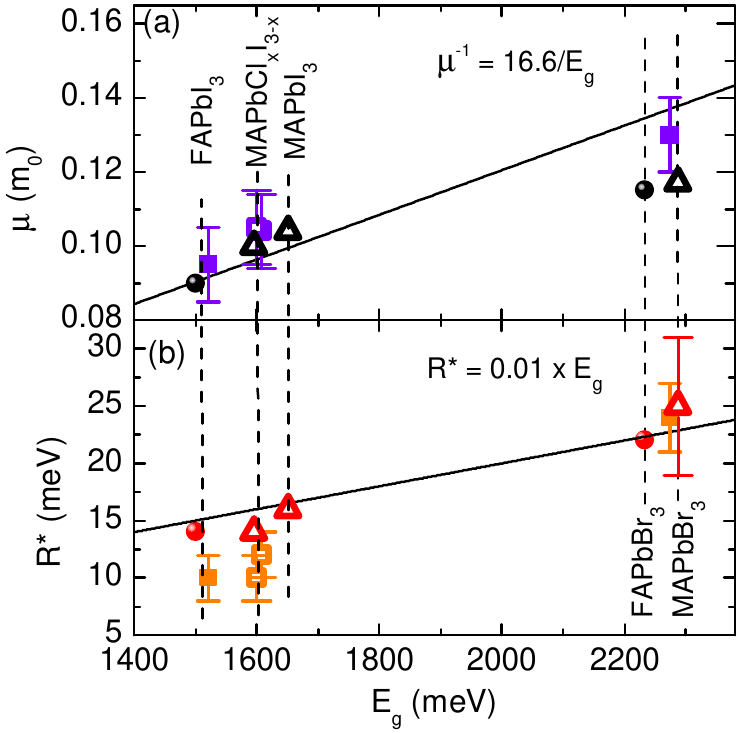}
\end{center}
\caption{(a), (b) Effective mass and binding energy as a function
of the energy gap respectively. Closed  (open) symbols mark
results for FA (MA) based families at 2K. Squares show values
deduced in the higher temperature tetragonal phase measured in the
temperature ranges given in Table ~\ref{Tab_t1}. The black solid
lines show a linear fit to the data.}\label{Fig4}
\end{figure}

The exciton effective masses are in good agreement with recent
predictions ($\mu$=0.09 -0.1 $m_0$) of density functional theory
for the tri-iodides where the theory has been adjusted to fit the
experimental values of the band gap\cite{Umari14,Menendez14}. Fig
\ref{Fig4} and Table \ref{Tab_t1} show that at 2K both $\mu$ and
$R^*$ are weakly increasing functions of the material band gap.
The relatively simple band structure of these materials, which
have non-degenerate band edges which are relatively isotropic and
have similar effective masses suggests that they can be described
by a simple semi-empirical two band $\textbf{k.p}$ Hamiltonian
approach\cite{Even15} with effective masses for the electron and
hole given by

\begin{equation}\label{kp}
\frac{1}{m_{e,h}} = \frac{1}{m_{0}}|(1 \pm
\frac{2m_{0}|P|^{2}}{E_{g}})|
\end{equation}

 where the Kane energy is $2m_{0}(|P|^{2})$,
 with $P=<\psi_{VB}|i\hbar\frac{\partial}{\partial x}|\psi_{CB}>$
 which is the momentum matrix element coupling electron wavefunctions in the conduction and
 valance bands, which is taken to be isotropic. The valence band consists mostly of Iodine p-like states and the
 conduction band is a mixture of hybridized Lead s and p-like
 states\cite{Even12,Umari14,Zhou14} in a similar form to III-V
 semiconductors. The value of the excitonic reduced mass is therefore

\begin{equation}\label{mred}
\frac{1}{\mu} = |\frac{1}{m_{e}}| + |\frac{1}{m_{h}}|
=\frac{4m_{0}|P|^{2}}{E_{g}}
\end{equation}

 Direct calculations of the Kane energy predict it to
 be in the range 5.3 - 6.3 eV for these materials \cite{Even14,Fang15}
 although recent theoretical mass values suggest a higher value\cite{Umari14,Menendez14}.
 Fig \ref{Fig4} shows that the reduced mass is quite well
 described using the band gap obtained from Eq.~\ref{mred}
 and a somewhat larger value of 8.3 eV for the Kane energy which fits
 all members of the family studied to within $10\%$. There is no significant
 difference between the material made using different organic
 cations (MA or FA) other than that associated with the small changes in band gap
 probably associated with changes in lattice parameter.

 The effective Rydberg is somewhat more material dependent, varying
 from 14 - 25 meV approximately as:

\begin{equation}\label{Rstar}
R^{*}=0.010*E_{g}
\end{equation}

 with values approximately 3 times smaller than previously reported from
 magneto-optical studies where only the 1s state was observed\cite{Tanaka03},
 but in good agreement with the value of 15 meV deduced by Even et
 al\cite{Even14} by fitting a much more highly resolved zero field exciton absorption
 spectrum.
 We attribute this difference to the much poorer resolution
 of the excitonic transitions in the previous magneto-optical studies and to the
 analysis of only a single excitonic transition. The low temperature values deduced
 here are all less than or comparable to the thermal
 energy at room temperature and suggest therefore that the entire
 family of organic-inorganic tri-halide perovskites can be
 expected to show rapid excitonic ionization at 300K. This conclusion
 is strengthened by the temperature dependent reduction of the exciton binding energy discussed below.

 We can use the conventional formula for the effective Rydberg to
 define an effective dielectric constant $\epsilon_{eff}$ which
 takes account of the screening of the Coulomb potential by the
 lattice. This value, in the range 7.5-9.8, is intermediate between the low ($\epsilon_{0}$ = 25.7)\cite{Wehrenfennig14} and high ($\epsilon_{\infty}$ = 5.6) frequency values \cite{Umari14,
 Brivio14}and decreases slightly with the increasing values of $\mu$
 and $R^{*}$ probably due to the  reduction in size of the exciton Bohr radius and differences in the crystal structure. This behaviour is characteristic of highly polar
 materials with strong electron-phonon coupling.  This was first
 discussed by Pollman and Buttner \cite{Pollman75} and by Kane
 \cite{Kane78} who showed that in II-VI and I-VII materials
 $\epsilon_{eff}$ was intermediate between the two values $\epsilon_{0}$ and
 $\epsilon_{\infty}$. In TlCl for example, a material with comparable parameters
  ($\epsilon_{\infty}$ = 5.4, $\epsilon_{0}$ = 37.6, $\mu$=0.18$m_{0}$) it was found that
   $\epsilon_{eff}$=2.48$\epsilon_{\infty}$.  In perovskites where there are
   multiple polar phonon modes, including rotational motion of the MA cations \cite{Wehrenfennig14,Umari14,
 Brivio14,Even14} such a calculation is more difficult but has recently
 been reported \cite{Menendez15} giving results consistent with the proposal that the value
 of $\epsilon_{eff}$ will again be intermediate between $\epsilon_{0}$ and
 $\epsilon_{\infty}$. It is noticeable that the values for
 $\epsilon_{eff}$ are significantly lower for the tri-bromides,
 where the higher reduced masses cause the binding energy to be
 larger, with a consequent reduction in the strength of the
 screening. Given the higher uncertainty in the
 values of R* for the MAPbBr$_{3}$, however, we do not consider the
 difference in $\epsilon_{eff}$ between the MA and FA tri-bromides to
 be significant at present.

 The values for $g_{eff}$ are slightly larger than reported
 previously \cite{Tanaka03} and are consistent with a linear spin
 splitting and show no evidence for the predicted Rashba splitting
 which it has been suggested may occur for some crystal structures
 in the high temperature phase \cite{Kim14,Even15}.

In order to be useful for understanding the properties of devices
operating at room temperature it is important to expand the
magneto-optical studies to as high a temperature as possible, in
particular as all the family of organic-inorganic tri-halide
 perovskites show a series of phase transitions at higher temperatures
 to more symmetric structures. The tri-iodides transform from an
  orthorhombic to a tetragonal structure with increasing temperature at around 150K accompanied by a
  small decrease in band gap.  The tri-bromides
  are cubic at room temperature\cite{Eperon14, Noh13}, and transform to tetragonal
  in the alloy family FAPbBr$_{3-x}$I$_{x}$ at around $x$=0.6, although when
  plotted as a function of the pseudocubic lattice parameter $a^{*}$ the
  band gap is a continuous function across the phase transition\cite{Eperon14}.
  As a function of temperature the tri-bromides transform to a
  tetragonal structure at around 240K and then become orthorhombic below
  150K\cite{Onada90}. The transition from the cubic to the tetragonal structure
  appears to have no significant influence on the band gap
  as measured either by absorption (SI Fig 3) or reflectivity \cite{Kunugita15},
  but the transition to a tetragonal structure at low temperatures
  produces a small shift in band gap of around 10 meV.

\begin{figure*}[]
\begin{center}
\includegraphics[width=1.3\columnwidth]{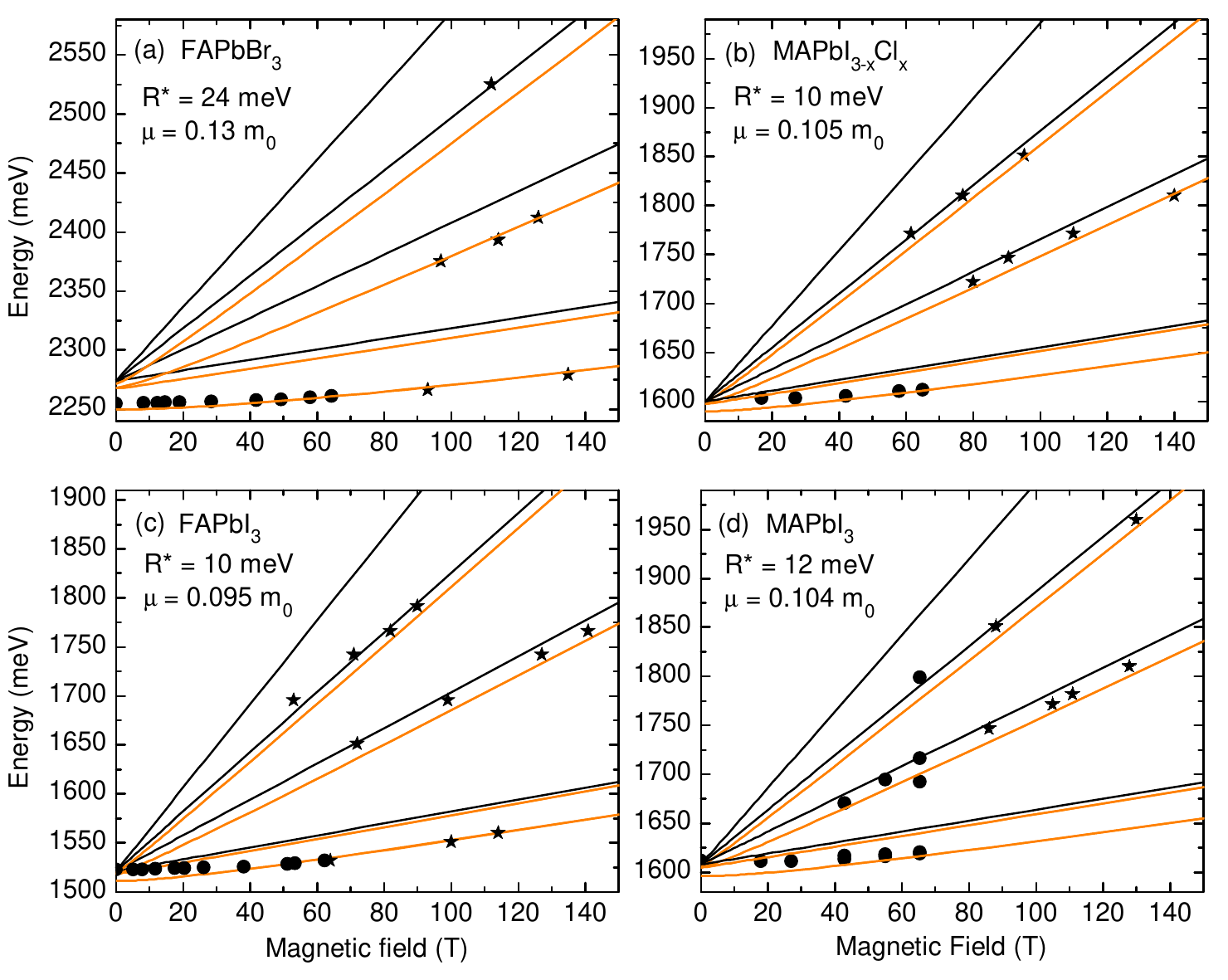}
\end{center}
\caption{(a)-(c) transition energies obtained from the
experimental data and results of the fit to the data for the high
temperature phases of MAPbI$_{3-x}$Cl$_{x}$, FAPbI$_{3}$, and
FAPbBr$_{3}$ respectively. The circles are data obtained using
long pulsed fields and the stars are data obtained during the
short pulsed field measurements both using unpolarized light.
Polarization was not used. The data from Miyata et
al.\cite{Miyata15} for the high temperature phase of MAPbI$_{3}$
are shown (d) for comparison. The results of the theoretical fit
are shown by grey and orange lines. Grey lines correspond to
interband transitions between Landau levels in the valence and
conduction bands. The orange lines show the strongly bound levels
of the hydrogen - like exciton.}\label{HighTfans}
\end{figure*}

  We have repeated the magneto-optical study at higher temperatures
  (140-200K), sufficiently high to be in the higher temperature crystallographic
  phase for each material as determined from the absorption measurements (SI Fig. 3). The
  absorption spectra are shown in the supplementary information (SI
  Fig. 1, Fig. 2) and the fan diagrams in Fig\ref{HighTfans}.  These
  are again fitted to give the band parameters for the high temperature phase shown
  in Fig~\ref{Fig4} and Table \ref{Tab_t1}. The experimental
  values deduced for the high temperature results have
  significantly greater statistical and systematic errors as fewer
  transitions are seen and these are less well resolved.  In particular the non-observation
  of a 2s state in the intermediate field range of 10 - 65 T,
  where it was strongest at low temperatures, makes the
  measurement of the exciton binding energy significantly less certain. An additional
  factor also is that it has been shown that \cite{Behnke78} the phonon contribution
  to the dielectric screening is reduced by the magnetic field due to
  the increasing exciton binding energy and so the values of R*
  deduced from the resonances observed at over 50T will be an
  overestimate of the zero field value as discussed previously by Miyata et
  al\cite{Miyata15}.

 In the the MAPbI$_{3}$ it was found\cite{Miyata15}that
 the effective mass in the higher temperature tetragonal phase was essentially the
 same as for the low temperature phase at a similar band gap. Our results here
  suggest that although the values in the high temperature phase
  may be a few $\%$ larger than at 2K this is at the limit of the
  accuracy of the high temperature fits and that essentially the
 effective masses of the whole family of organic-inorganic tri-halide
 perovskites can be reasonably described in terms of the single
 parameter of the band gap through Eq.~\ref{mred} irrespective of
 the crystal phase or temperature.


 The phase transition to a tetragonal phase in the region of 150K is
 likely to be more significant for the dielectric constant and consequently the exciton binding
 energy,
  as additional collective rotational motion of the organic cations
 becomes allowed. This has been shown to cause an increase in the dielectric
 constant \cite{Poglitsch87}, which led Even et al \cite{Even14}
 to propose that this could cause a decrease in binding
 energy to values of the order of 5 meV\cite{Even14} which they again
 deduced from fitting the excitonic absorption. Using a similar
 type of analysis Yamada et al\cite{Yamada15} have suggested that
 there is a continuous decrease in binding energy from 30 meV at
 low temperature to 6 meV at 300K. The existence of a decreased
 exciton binding energy in the high temperature phase
 was supported by our previous high temperature magneto-optical
 measurements\cite{Miyata15}. For all the tri-iodide materials analysed here
 the resonant transitions are only observed at very high fields (above
 50T) but even so the exciton binding energies are reduced
 to values of order 10$\pm$3 meV with a corresponding increase in $\epsilon_{eff}$. In addition the anomalously low diamagnetic
 shift of the 1s exciton state suggests that there is a field dependent
 increase in the 1s binding energy\cite{Behnke78} which leads us to
 conclude that the low field exciton binding energy
  has fallen to values of order 5 meV or less.  By
 contrast the FAPbBr$_{3}$ shows no significant reduction in the
 exciton binding energy in the higher temperature cubic state,
 probably because the original binding energy corresponds to a
 frequency which is too large to respond to the additional screening
 and consequently the low field diamagnetic shift is close to the theoretical prediction.

\section{Conclusions}

In conclusion our results provide a general relation for the low
temperature exciton binding energy (\ref{Rstar}) and reduced
effective mass (\ref{mred}), which are determined principally by
the material band gap $E_g$ for the whole organic-inorganic lead
tri-halide perovskite family. The exact nature of the organic
cations (MA or FA) is shown to have relatively little influence,
producing only small changes in effective mass which are
correlated with the changes in band gap. This suggests that simple
two band \textbf{k.p} perturbation theory should enable accurate
modelling of the properties of devices made from materials in this
system. Our results also demonstrate that the low temperature
values of the exciton binding energies are small compared to the
thermal energy at 300K which makes a significant contribution to
understanding the excellent performance of these devices. For the
tri-iodides these values fall still further in the high
temperature phase of the materials probably due to increased
screening but this is not observed in the FA tri-bromides.

\section{Experimental Method}

Perovskite precursor synthesis: Formamidinium iodide (FAI) and
formamidinium bromide (FABr) were synthesised by dissolving
formamidinium acetate powder in a 1.5$x$ molar excess of 57$\%$
w/w hydroiodic acid (for FAI) or 48$\%$ w/w hydrobromic acid (for
FABr). After addition of acid the solution was left stirring for
10 minutes at 50$^{0}$ C.  Upon drying at 100$^{o}$C, a
yellow-white powder is formed. This was then washed twice with
diethyl ether and recrystallized with ethanol, to form white
needle-like crystals. Before use, it was dried overnight in a
vacuum oven. To form FAPbI$_{3}$ and FAPbBr$_{3}$ precursor
solutions, FAI and PbI$_{2}$ or FABr and PbBr$_{2}$ were dissolved
in anhydrous N,N-dimethylformamide (DMF) in a 1:1 molar ratio, at
0.55M of each reagent, to give a 0.55M perovskite precursor
solution. To form CH$_{3}$NH$_{3}$PbI$_{3-x}Cl_{x}$ precursor
solutions, Methylammonium iodide (MAI) and lead chloride
(PbCl$_2$) were dissolved in a 40$\%$ w/w DMF solution in a 3:1
molar ratio. For MAPbI$_{3}$, precursor solutions were prepared
separately by dissolving lead iodide (PbI$_2$) in DMF (450 mg/ml),
and MAI in isopropanol (50 mg/ml), respectively.

 Film formation: All the samples were prepared in a nitrogen-filled
glovebox on glass substrates cleaned sequentially in hallmanex,
acetone, isopropanol and O$_{2}$ plasma. Immediately prior to the
Formamidinium (FA) sample film formation, small amounts of acid
were added to the precursor solutions to enhance the solubility of
the precursors and allow smooth and uniform film formation.
38$\mu$l of hydroiodic acid (57$\%$ w/w) was added to 1ml of the
0.55M FAPbI$_{3}$ precursor solution, and 32$\mu$l of hydrobromic
acid (48$\%$ w/w) was added to 1ml of the 0.55M FAPbBr$_{3}$
precursor solution. FA Films were then spin-coated from the
precursor plus acid solution on warm (85$^{o}$C) substrates for
45s at 2000rpm, followed by annealing at 170$^{o}$C in air for 10
minutes. This gave very uniform pinhole-free layers, ~350nm thick,
of FAPbI$_{3}$ or FAPbBr$_{3}$.

The Methylammonium (MA) samples were prepared following methods
described previously \cite{Lee12, Heo13}. In brief,
CH$_{3}$NH$_{3}$PbI$_{3-x}Cl_{x}$ films were prepared by
spin-coating for 60s at 2000 rpm on warm (85$^{o}$C) substrates,
followed by annealing at 100$^{o}$C in air for 1 hour. The
CH$_{3}$NH$_{3}$PbI$_{3}$ films were prepared from a PbI$_2$ layer
which was first deposited on cleaned glass by spin-coating at
6000rpm for 30s from a precursor solution, followed by drying at
70$^{o}$C for 5 min. Then the MAI layer was deposited on the dried
PbI2 layer by spin-coating at 6000rpm for 30s from a precursor
solution, followed by annealing at 100$^{o}$C for 1 hour. For the
MAPbBr$_{3}$ films lead(II) acetate trihydrate (Sigma) and MABr
(Dyesol) were dissolved in anhydrous N,N-Dimethylformamide (DMF)
at a 3:1 molar ratio with a final perovskite precursor
concentration of ~40 wt$\%$. Then, the solution was spin-coated at
2000 rpm (for 45 s) in a nitrogen-filled glovebox. After
spin-coating, films were left to dry at room temperature in the
glovebox for 5-10 minutes and then annealed at 100$^{o}$C for 15
minutes.
 The perovskite films were all sealed by spin-coating a layer
of the insulating polymer poly(methyl methacrylate) (PMMA) at
1000rpm for = 60s (precursor solution 10mg/ml in cholorobenzene)
on top in order to ensure air-and moisture-insensitivity.

The magneto-transmission measurements were performed in pulsed
magnetic fields $\leq 70$~T ($\simeq 500$~ms) in LNCMI Toulouse. A
tungsten halogen lamp provides a broad spectrum in the visible and
near infra-red range and the absorption was measured in the Faraday
configuration with the $c$-axis of the sample parallel to magnetic
field. The circular polarization was introduced in-situ. Rotation
between $\sigma^{+}$ and $\sigma^{-}$ polarized light was done by
change of the direction of the magnetic field. A nitrogen cooled CCD
detector analyzed the transmitted light dispersed by a spectrometer.
Differential transmission spectra were produced by normalizing all
the acquired spectra by the zero field transmission or by signal of
the halogen lamp. The sample was immersed in liquid helium cryostat.
The measurements have been performed either in superfluid or in gas
helium. Measurements to higher fields $\leq 150$~T were performed
using a semi-destructive technique and pulse lengths of $\simeq
10$~$\mu$s, where transmission of series of laser diodes and
Ti-Sapphire tunable laser through the sample was measured by fast
(100MHZ) silicon detector and high speed digital
oscilloscope\cite{Robin13,Miyata15}. The sample was mounted inside a
non conducting helium flow cryostat and was cooled separately for
each measurement. The magnetic fields were generated by a
semi-destructive single turn coil system using 10mm
coils\cite{Robin13,Portugal99}

\section{Acknowledgements}

This work was partially supported by  ANR JCJC project milliPICS,
the Region Midi-Pyr\'en\'ees under contract MESR 13053031. We also
acknowledge support by EuroMagNET II under the EU Contract Number
228043. The authors also thank: Meso-superstructured Hybrid Solar
Cells -MESO NMP-2013-SMALL7-604032 project, the Engineering and
Physical Sciences Research Council (EPSRC), the European Research
Council (ERC-StG 2011 HYPER Project no. 279881).





\bibliographystyle{rsc} 

\end{document}